\documentclass[10pt,letterpaper]{article}
\usepackage{cite}
\usepackage{OE}
\usepackage{amsmath}
\usepackage{bm}

\usepackage{graphicx}
\usepackage{hyperref} %% optional

\usepackage{xspace}
\usepackage{units}

\newcommand{\oam} {orbital angular momentum\xspace}

\newcommand{\KEEP}[1]{#1}{}

\makeatletter
\renewcommand\@biblabel[1]{#1.}
\makeatother

\begin{document}

\title{Bright and dark helices of light}

\author{Ole Steuernagel}

\address{School of Physics, Astronomy and Mathematics,
University of Hertfordshire, Hatfield, AL10 9AB, UK}

\email{O.Steuernagel@herts.ac.uk}

\date{\today}

%%%%%%%%%%%%%%%%% END OF PREAMBLE %%%%%%%%%%%%%%%%
\begin{abstract}
  Laser beams can be made to form bright and dark intensity helices of
  light. Such helices have a pitch length on the order of a wavelength
  and may have applications in lithography and the manipulation of
  particles through optical forces. The formation of bright helices is
  more strongly constrained by optical resolution limits than that of
  dark helices, corresponding scaling laws are derived and their
  relevance for photo-lithography pointed out. It is shown how to
  arrange dark helices on a grid in massively parallel fashion in
  order to create handed materials using photo-lithographic
  techniques.
\end{abstract}


%                           NOTE: \ocis{} IS ALIASED TO \pacs{} BUT MUST
%                           FORMAT THE TERMS CORRECTLY FOR EACH JOURNAL

% \pacs{
% 42.25.-p, %Wave optics
% 42.25.Hz, %Interference
% 42.60.Jf, %Beam characteristics: profile, intensity, and power; spatial pattern formation
% 42.82.Cr, %Fabrication techniques; lithography, pattern transfer
% 37.10.Vz  %Mechanical effects of light on atoms, molecules, electrons, and ions
% }

%\maketitle

\bibliographystyle{osajnl}

%\input{Helix.bbl}
%\bibliography{Helix_bibliography}

\section{Introduction}

In recent years there has been considerable interest in monochromatic
light beams, such as Laguerre-Gauss beams~\cite{Allen_PRA92}, carrying
orbital angular momentum. These are known to have helical wave
fronts~\cite{Allen_PRA92,Padgett_CP00,Padgett_PT04} which have been
investigated using interferometric
techniques~\cite{Harris_OC94,Vaughan_JOPA99,Leach_OE06}. Several
approaches that create helical beams using the action of Gouy's
phase~\cite{Ole_AJP05} have been suggested~\cite{Lekner_JOPA04} and
experimentally implemented~\cite{Hamazaki_OE06,Baumann_OE09},
combinations of plane waves have been used~\cite{Becker_OE11}, whilst
the use of Bessel-beams has also been
suggested~\cite{Volke-SepulvedaJPB09}.

Few setups have been envisaged using counterpropagating monochromatic
beams, although only these yield helices with pitch lengths on the
order of their wavelength. Notable exceptions studying bright helices
are~\cite{Volke-SepulvedaJPB09,Staliunas_cond.mat.99,Bhattacharya_OC07}. Here,
dark helices are investigated for the first time; they are not
resolution limited and therefore provide us with better contrast than
bright helices.

\KEEP{In section~\ref{sec_2_interference}} bright and dark intensity
helices of light are introduced in terms of screw
dislocations~\cite{Nye_Berry_PRSLA74} of ordinary standing waves. This
allows us to quantify some of their basic characteristics. In
section~\ref{sec_4_Single_DarkHelices} it is shown that using
superpositions with unequal weight to generate the intensity helices
allows us to control their widths. Section~\ref{sec_dark_scaling}
describes the scaling of the electric field strength around dark and
bright intensity helices and how this limits the resolution of bright
helices compared to dark ones.
Section~\ref{sec_4_LatticesOfDarkHelices} shows how both types can be
arranged in large arrays in massively parallel fashion making them
potentially suitable for the creation of helical metamaterial using
bulk lithography. Section~\ref{sec_5_ApplicationsConclOutlook}
concludes with a survey of applications and the outlook.

\section{Helices through interference~\label{sec_2_interference}}

\begin{figure}[ht]
\begin{center}
  \begin{minipage}[b]{0.29\linewidth} %  "b" to out the captions on the same line
     \includegraphics[width=0.75\linewidth,height=0.85\linewidth]{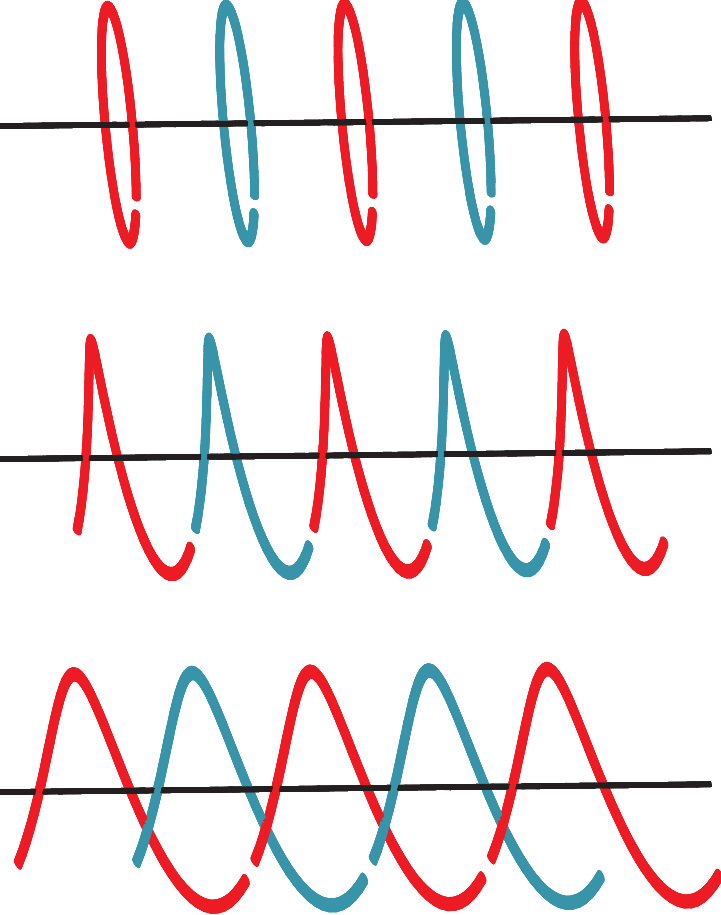}
     \put(-80,99){\rotatebox{0}{\mbox{(a)}}}
  \end{minipage}   % no empty line here so the figures are side by side
  \hspace{0.0035\linewidth}
  \begin{minipage}[b]{0.33\linewidth} %  "b" to out the captions on the same line
     \includegraphics[width=0.995\linewidth,height=0.85\linewidth]{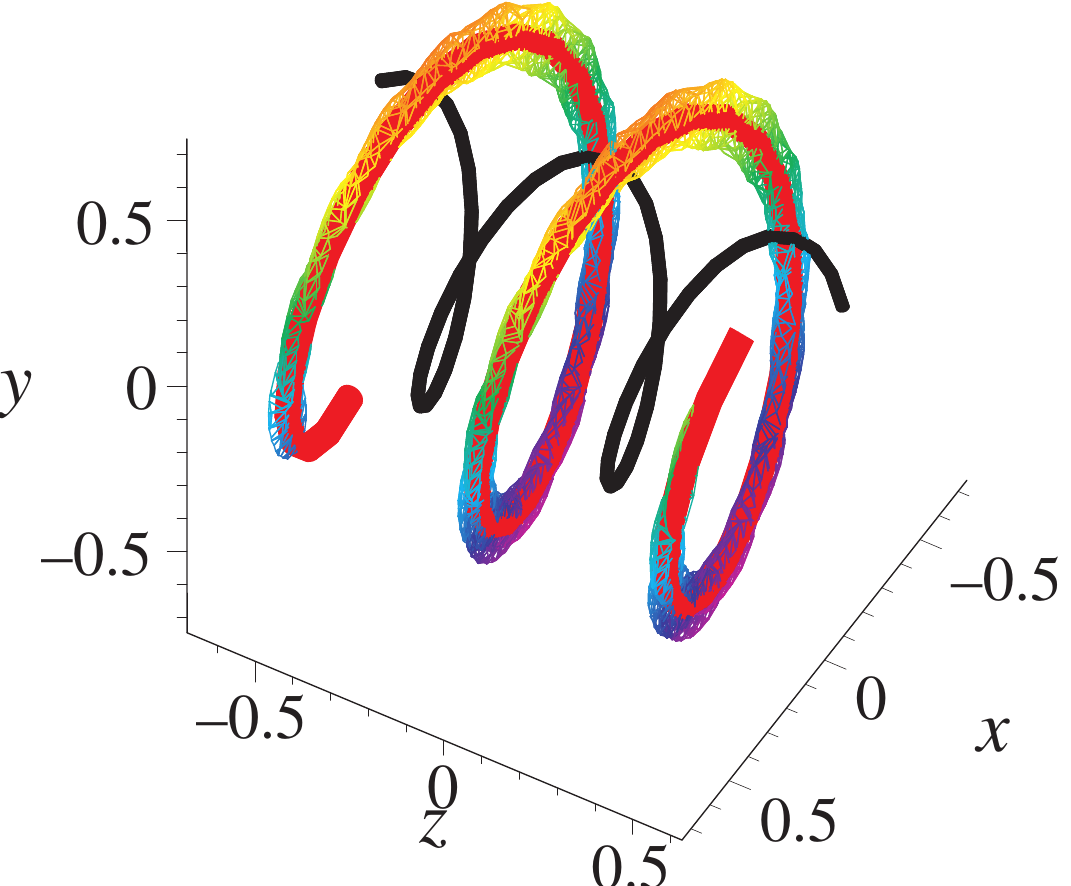}
     \put(-120,99){\rotatebox{0}{\mbox{(b)}}}
  \end{minipage}   % no empty line here so the figures are side by side
  \hspace{0.0035\linewidth}
  \begin{minipage}[b]{0.33\linewidth}
     \includegraphics[width=0.998\linewidth,height=0.75\linewidth]{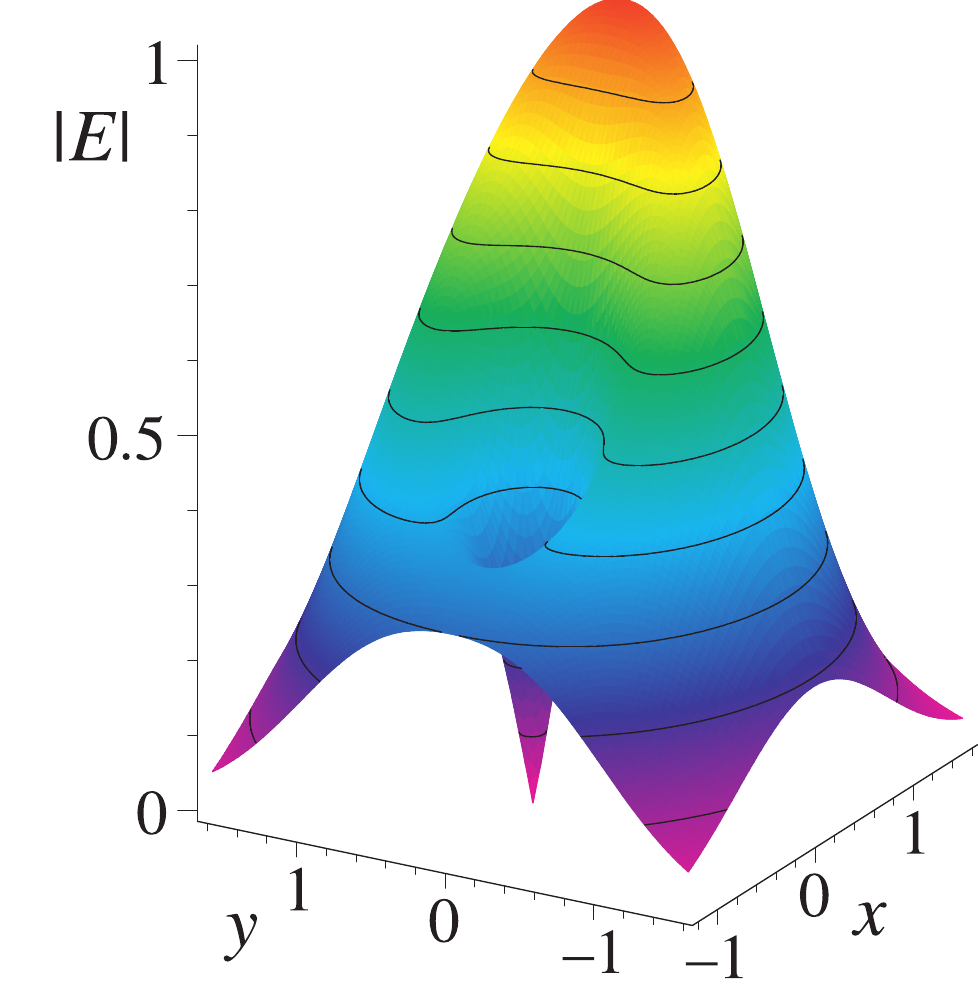}
     \put(-120,99){\rotatebox{0}{\mbox{(c)}}}
  \end{minipage}
\end{center}
\caption{{(a)} illustrates the `pancake' scenario explaining that
  helices form with pitch heights which are integer multiples of
  $\lambda/2$. {(b)} shows the superposition $\frac{1}{2}
  u_{0,0}(x,y,z) + u_{0,1}(x,y,-z)$ yielding a single bright helix
  (red line) enveloping a single dark helix (black line) with the
  minimal pitch length $\lambda/2$ [colored mesh was numerically
  determined as the locations with 90\% of peak intensity]. {(c)}
  shows the magnitude~$|E|$ of its focal field distribution
  illustrating quadratic variation of $|E|$ around the maximum and
  linear variation around the minimum. In {(b)} and {(c)} the $x$- and
  $y$-axes are given in units of focal beam radius~$w_0$, $z$-axis in
  units of $\lambda$, in {(c)} the peak intensity is normalized to
  unity.}
\label{Fig_1_single_helix}
\end{figure}

\begin{figure}[ht]
\begin{center}
  \begin{minipage}[b]{0.47\linewidth}
    \includegraphics[width=0.7\linewidth,height=0.7\linewidth]{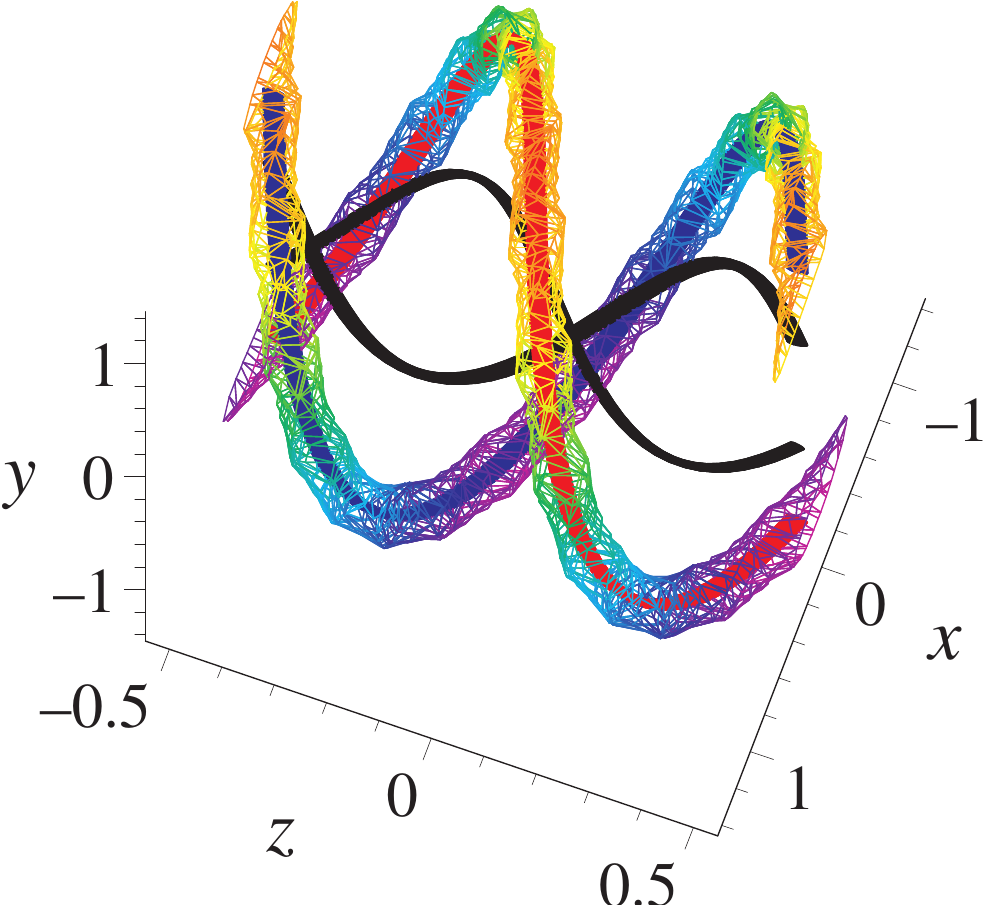}
     \put(-110,105){\rotatebox{0}{\mbox{(a)}}}
  \end{minipage}   % no empty line here so the figures are side by side
  \hspace{0.035\linewidth}
  \begin{minipage}[b]{0.47\linewidth}
    \includegraphics[width=0.7\linewidth,height=0.7\linewidth]{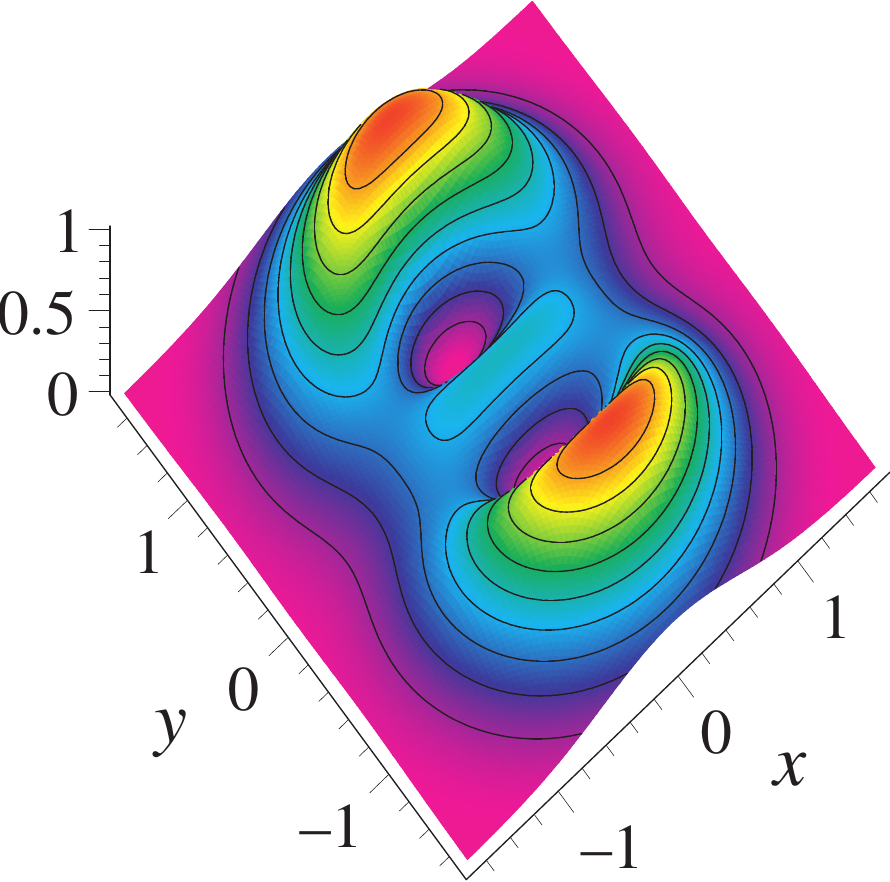}
     \put(-110,105){\rotatebox{0}{\mbox{(b)}}}
     \put(-125,90){\rotatebox{0}{\mbox{$I$}}}
  \end{minipage}
\end{center}
\caption{  Superposition $u_{0,2}(x,y,z) + \frac{1}{2}
u_{1,0}(x,y,-z)$
  yields a pair of dark and bright helices with
  pitch $\lambda$ {(a)}; its focal
  intensity distribution {(b)} [all units as in
  Fig.~\ref{Fig_1_single_helix},
  note that compared to Fig.~\ref{Fig_1_single_helix} the helices'
  orientation is reversed].}
\label{Fig_2_double_helix}
\end{figure}

Bright and dark interference fringes extending across a standing
wave's cross-section are ubiquitous. In optics they have found
applications in Lippmann's photo\-gra\-phy and Gabor's
holo\-graphy. They arise in laser cavities and interferometry and are
used as transporters~\cite{Kuhr_SCI95} and imaging
elements~\cite{Meschede_Metcalf_03} in atom optics. They are on the
order of half a wavelength $\lambda$ of the interfering light apart
and since light beams are more than a wavelength wide and can overlap
over considerable areas these bright fringes often resemble very tall
stacks of pancakes~\cite{Kuhr_SCI95}. They can also be slightly
modified (the pancakes become deformed), due to the dispersive effects
of Gouy's phase, in interfering multimode
beams~\cite{Ole_AJP05}. Instead of using the superposition of
identical beams we want to consider the interference between two
monochromatic, collinear, \emph{counterpropagating} partial waves with
{\emph{different}} orbital angular
momenta~\cite{Allen_PRA92,Padgett_CP00}.  In this case one can imagine
the pancakes (which are centered on the beam axis,
compare~Fig.~\ref{Fig_1_single_helix}~{(a)} top) to be cut open on one
side, then splayed open and glued to the opposing cut faces of their
neighbours, compare~Fig.~\ref{Fig_1_single_helix}~{(a)} middle. In the
most tightly wound case such a screw
dislocation~\cite{Nye_Berry_PRSLA74} yields a helix with pitch length
$z_1=\lambda/2$, see Fig.~\ref{Fig_1_single_helix}.
The common pancake stack scenario can be viewed as the special case of
a degenerate helix with orbital angular momentum difference $l-l'=0$
and associated pitch length $z_0=0$. Alternatively to linking up
nearest neighbours, the cut-open pancakes can be connected to their
second nearest neighbour with pitch length~$z_2=\lambda$ leaving the
nearest neighbour to form part of a second helix on the opposite side
of the beam axis, compare~Fig.~\ref{Fig_1_single_helix}~{(a)}
bottom and Fig.~\ref{Fig_2_double_helix}.

In general the helices' pitch lengths~$z_{l-l'}$ (orbital angular
momentum quantization implies that $l$ is an
integer~\cite{Allen_PRA92}) are determined by the difference $l-l'$ in
orbital angular momentum of the used laser beams and
obey~$z_{l-l'}=\lambda (l-l')/2$. A negative helix length~$z_{l-l'}$
describes inverted handedness. Different superpositions can yield a
larger number of intertwined helices and concentric shells of helices,
compare Fig.~\ref{Fig_2_double_helix}.  Per nodal shell, there tend to
exist a number of $|l-l'|$ separate bright helices with dark helically
wound regions between them, see
references~\cite{Padgett_CP00,Padgett_PT04} and text following
eq.~(\ref{eq_LG_modes}) below. Details depend on the mode structures
and relative weights of the employed superpositions\KEEP{}{, see
  sections~\ref{sec_2_interference}
  and~\ref{sec_dark_scaling}}.

The helices' widths can be controlled through the width of the laser
beams' waists~$2 w_0$. Their lengths are practically limited by the
laser beams' divergence. If the beams' cross-sections are elliptically
deformed, so are the helices. Suitable laser beam modes that carry
\oam, such as Laguerre-Gauss--modes~\cite{Allen_PRA92}, are readily
available~\cite{Padgett_CP00,Padgett_PT04,Harris_OC94,Vaughan_JOPA99,Marrucci_PRL06}
and the required superpositions are straightforward to
implement~\cite{Leach_NAT04}.
The relative phases of the interfering beams can be shifted such that
the intensity-helices revolve around their main axis (including very
rapid rotations through frequency detuning between the two opposing
beams~\cite{MacDonald_SCI02,Marrucci_PRL06}).

For specificity, we will consider the paraxial Laguerre-Gauss modes
$U_{p,l}(x,y,z)$~\cite{Allen_PRA92,Pampaloni_2004} (formed from
`generalized Laguerre polynomials'~$L_{p,|l|}$ rather than, say, orbital
angular momentum carrying Bessel-beams~\cite{Volke-SepulvedaJPB09})
\begin{eqnarray}
 & U_{p,l} & \! \! \! \! \left(  z_R,\lambda; r,\phi,z \right) = \sqrt
{{\frac { 2}{(1+\delta_{0,l}) \pi} \frac{ p!}{ \left( p+|l| \right)
!}}}
\nonumber \\
& \times &
 \left( {\frac {\sqrt {2}r}{w \left( z\right) }} \right)
^{|l|}
 L_{p,|l|} \left(2\,{
\frac {{r}^{2}}{ w \left( z \right)^2 }} \right)
 \left( \frac{e^{i l\phi}}{w \left( z\right)} \right)
\nonumber \\
& \times &
 \exp \left[{-{\frac {{r}^{2}}{ w \left( z \right)^2}}}
 -i \left( 2\,p+|l|+1
 \right) \zeta(z)
 + {\frac {i k \,{r}^{2} }{2 \varrho(z)}}\right] \, .
\label{eq_LG_modes}
\end{eqnarray}

Here $\lambda$ is the laser's wavelength and $k=2\pi/\lambda$ its wave
number; $x$ and~$y$ are the transverse coordinates, and $z$
parameterizes the beam axis. Kronecker's $\delta$-function takes care
of the correct normalization. The azimuthal angle~$\phi$ is connected
to the transverse coordinates via the relation~$e^{i l\phi} =
(x+iy)^l$: $l$ is the integer orbital angular momentum index with the
associated orbital angular momentum $L_l=\hbar
l$~\cite{Allen_PRA92,Padgett_CP00}. The discrete radial index~$p$
counts the number of nodal rings in the radial
direction~$r=\sqrt{x^2+y^2}$. The beam radii are given by $w(z)=w_0
\sqrt{ 1+z^2/z_R^2}$ with the beam waist radius $w_0=\sqrt{\lambda z_R
  / \pi}$ where $z_R$ is the Rayleigh-length which also parameterizes
the Gouy-phase shifts~$\zeta(z)=\arctan(z/z_R)$ and the wavefront
radii~$\varrho(z)=(z^2+z_R^2)/z$~\cite{Allen_PRA92,Pampaloni_2004}.

In order to form bright (or dark) light helices we have to create
constructive (or destructive) interference along the entire phase
front. For this it suffices to interfere two radially matching pure
Laguerre-Gauss laser modes, with suitably chosen amplitudes~$C$
and~$C'$, travelling in opposite directions, namely, to form a
superposition
\begin{eqnarray}%{l}
\label{eq_Superposition_00_01}
\mathbf{E}(x,y,z;t) =  E(x,y,z;t) \cdot \mathbf{P} =  
[ C \, u_{p,l}(x,y,z) + C' u'_{p',l'}(x,y,-z) ]\cdot e^{-i \omega
  t} \cdot \mathbf{P} \, ,
\end{eqnarray}
where we used $\omega=c_0 k$ for the angular frequency of the light
and the paraxial solutions
\begin{eqnarray}
u_{p,l}(x,y,z)=U_{p,l}(x,y,z) e^{ikz} \; .
\label{eq_ParaxModes}
\end{eqnarray}
We assume that the laser is uniformly polarized and the wave front
curvatures can be neglected. Thus the electric field~$\mathbf{E}$ is a
scalar field multiplied with the polarization vector~$\mathbf{
  P}$~\cite{Haus.book}. From this superposition we can extract the
spatially fast-oscillating term~$e^{i(kz+l\phi)}$ which yields the
interference term $\cos(2 k z+ (l-l') \phi)$. When a full turn
($\Delta \phi = 2 \pi$) along the intensity helix is tracked the
ensuing spatial shift yields the pitch length in the $z$-direction
\begin{eqnarray}
z_{l-l'}=(l-l')2\pi/(2k) = \lambda(l-l')/2 \, .
\label{eq_PitchLengt}
\end{eqnarray}
This confirms the intuitive description in terms of a screw
dislocation given above. Note that the interference between two beams
travelling in the same direction yields interference
fringes~\cite{Harris_OC94} but not helical intensity spirals on the
wavelength scale; instead it forms intensity rods which are modulated
by Gouy's phase thus forming rods with a
twist~\cite{Paterson_SCI01,Lekner_JOPA04}, very similar to those
displayed in Fig.~1 of reference~\cite{Dholakia_PhysWorld02}.

Intuitively, this should not be too surprising since in a forward-only
configuration light following a helical path would travel
superluminally. The effects discussed here are not a consequence of
Gouy's phase, unlike the helical intensity distributions considered in
references~\cite{Lekner_JOPA04} and~\cite{MacDonald_OC02} or the nodal
lines investigated in reference~\cite{Leach_NAT04}, all giving rise to
`half-oscillations'~\cite{Ole_AJP05} on the scale of the beam's
Rayleigh length.

The effects discussed here necessarily require interference of two
counterpropagating partial waves; this implies that the medium needs
to be sufficiently transparent. Also note that reflection by a mirror
inverts the helicity of a mode and therefore beams carrying \oam,
say within a laser cavity, form intensity helices (but not single
standalone helices since $|l-l'|\neq 1$ in the reflection case).

\section{Some features of single dark helix
  beams~\label{sec_4_Single_DarkHelices}}

In this section some features of the single-helix superposition
\begin{eqnarray}
\Sigma(x,y,z;C) =  C \cdot
u_{0,0}(x,y,z) + u_{0,1}(x,y,-z) 
\label{eq_single_helix_Sigma}
\end{eqnarray}
of an ordinary Gaussian beam~$u_{0,0}$ with a counterpropagating
Laguerre-Gauss mode~$u_{0,1}$ with one unit of \oam are elucidated.

The single-helix superposition~$\Sigma(C)$ forms a single bright and a
single dark helix, the Gaussian beam's relative weight fixes the dark
helix' radius since the field's zero {in the focal plane} occurs
at~$(x_0,y_0,0)=(-C \, w_0/\sqrt{2},0,0)$. The stronger the field of
the added Gaussian beam the more the zero of the Laguerre-Gaussian
mode gets displaced from the beam axis, compare
Figs.~\ref{Fig_2_double_helix}
and~\ref{Fig_3_self-overlapping_double_helices}. For small values of
the coefficient $C$ the dark helices are slim and steep, yet, unlike
bright helices they remain distinguishable, see
Fig.~\ref{Fig_3_self-overlapping_double_helices}.

In general, determination of the precise form of helices along the
beam axis requires solving an implicit problem. For beams formed from
superpositions of the form~(\ref{eq_Superposition_00_01}) (of which
$\Sigma (C)$ of eq.~(\ref{eq_single_helix_Sigma}) is a special case) the
following mapping allows for a quick determination of the approximate
location of the helices.
%\begin{eqnarray}{r,c,l}
\begin{align}
  (x(z)+i y(z)) & = (x_0+i y_0)
  \sqrt{(1+z^2/z_R^2)}\exp(i\chi(z))
\label{eq_complex_plane_representation}  \\
  \mbox{with the phase }\quad \chi(z) & = -(2 (1+p+p')+ |l|+|l'|)
  \zeta(z) +\frac{k r^2}{\rho(z)} + 2 k z .
\end{align}
This is to be read as a complex-plane representation of a mapping of a
helix' focal plane location~$(x_0,y_0)$ to its location~$(x(z), y(z))$
along the beam axis.

\begin{figure}[!b]
\begin{center}
  \begin{minipage}[b]{0.47\linewidth}
     \includegraphics[width=0.7\linewidth,height=0.7\linewidth]{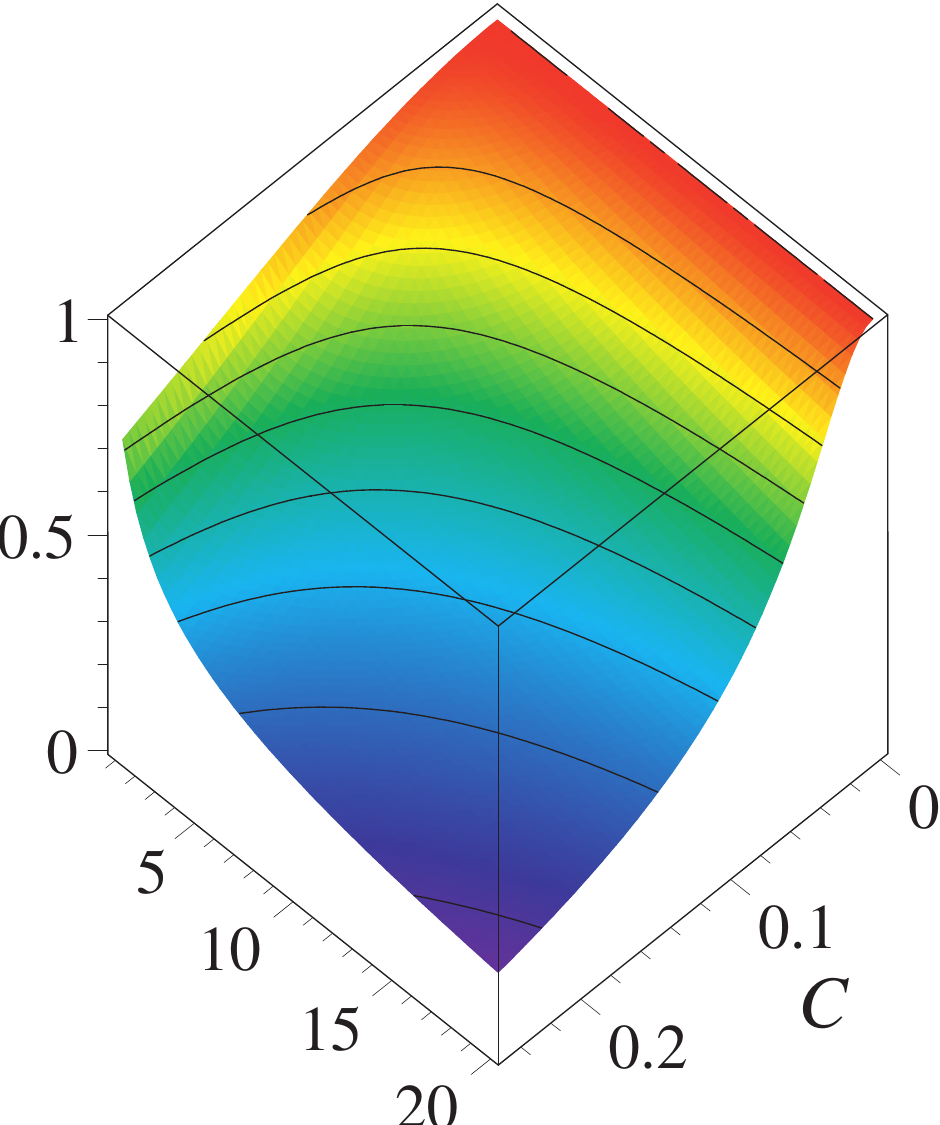}
  \end{minipage}   % no empty line here so the figures are side by side
     \put(-184,85){\rotatebox{0}{\mbox{A}}}
     \put(-170,10){\rotatebox{0}{\mbox{$z_R$}}}
\end{center}
\caption{ The frequency anisotropy~$A=\omega_x/\omega_\eta=|\partial_x
  E|/|\partial_\eta E|$ of the transversal trapping potential of dark
  helices of a single-helix beam~$\Sigma(C)$. The anisotropy $A <1$,
  $A\approx 1$ if one uses small helix radii ($C$ small);
  qualitatively the same behaviour was seen for bright helices in
  reference~\cite{Bhattacharya_OC07} [$z_R$ measured in units of
  $\lambda$].}
\label{Fig_Freq_Ratio}
\end{figure}

Next we determine the field gradients for a dark helix formed from a
single-helix beam $\Sigma(C)$ at the intensity zero in the focal plane
$(x_0,y_0,0)=(-C w_0/\sqrt{2},0,0)$
\begin{align}
  \partial_x E & = 2 \sqrt{\pi} \exp(- C^2/2) / (z_R \lambda),
 \label{eq_grad_x} \\
  \partial_y E & =  -i \partial_x E, \label{eq_grad_y} \\
  \mbox{and  \quad}
  \partial_z E & = C (-3\lambda+ C^2\lambda+4 z_R \pi)/\sqrt{ 2 \pi z_R
  \lambda} \cdot \partial_y E \; ; \label{eq_grad_z}
\end{align}
here, the complex phases represent the gradients' phase differences
within the optical cycle. The gradients can be used to derive the
exact pitch angle~$\alpha$ the helix forms with the focal plane
\begin{align}
  \alpha = \arctan \left( \frac{-\partial_y E}{\partial_z E} \right) =
  -\arctan \left( \frac{\sqrt{2\pi z_R \lambda}}{ C( -3\lambda+
      C^2\lambda+4 z_R \pi)} \right) \; .
\label{eq_pitch_angle}
\end{align} 
In this derivation the first step arises from the observation that
since the helix passes through $(-C \, w_0/\sqrt{2},0,0)$ its tangent
vector has the form $\boldsymbol{\tau} = (0,\tau_y,\tau_z)$. Since the
helix is dark it moreover fulfils the condition
$\tau_y\cdot \partial_y E + \tau_z\cdot \partial_z E =0 $ which
implies eq.~(\ref{eq_pitch_angle}) above. For large values of $C$ and
$z_R/\lambda$ this expression conforms with the `naive'
expectation~$|\alpha| \approx \arctan({h}/{x_0}) = \arctan(\lambda /
(2^{3/2} \pi C w_0))$.

With the abbreviation $g=\partial_z E/\partial_y E$, we can write down
the normalized focal tangent vector~$\boldsymbol{\tau} =
(1+g^2)^{-1/2}(0,-g,1)$ and complete the associated Frenet trihedron
by forming the third vector~$\boldsymbol \eta=\mathbf{\hat x} \times
\boldsymbol \tau = -(1+g^2)^{-1/2}(0,1,g)$.

A dark helix can be used as the core of a
blue-detuned~\cite{Chu_RMP98}, helical atom
waveguide~\cite{Bhattacharya_OC07}. Whereas~$\boldsymbol{\tau}$ points
in the direction of free movement, $\mathbf{\hat x}=(1,0,0)$
and~$\boldsymbol{\eta}$ form the principal axes of the local
transverse trapping potential
\begin{align}
  V(x,\eta) & \propto |\partial_x E|^2
  x^2+ |\partial_\eta E|^2 \eta^2\; , \qquad \mbox{ where } \quad 
  \partial_\eta E = \boldsymbol{\eta} \boldsymbol{\cdot \nabla} (E) \qquad \mbox{ with }\\
  \partial_\eta E & = -i \partial_xE \cdot \sqrt{\frac{ 2 z_R \pi + 9
      C^2 \lambda-6 C^4 \lambda -24 C^2 \pi z_R+C^6 \lambda + 8 C^4
      \pi z_R+16 C^2 \pi^2 z_R^2/\lambda}{2 z_R \pi}}
\label{eq_trap_potential}
\end{align}
This can be used to determine the transverse potential's trapping
frequencies and their anisotropy
ratio~$A=\omega_x/\omega_\eta$~\cite{Bhattacharya_OC07}, which is plotted 
in Fig.~\ref{Fig_Freq_Ratio}.

\section{Dark helices are more sharply contoured than bright
  helices~\label{sec_dark_scaling}}

\begin{figure}[t]
\begin{center}
  \begin{minipage}[b]{0.47\linewidth}
     \includegraphics[width=0.7\linewidth,height=0.7\linewidth]{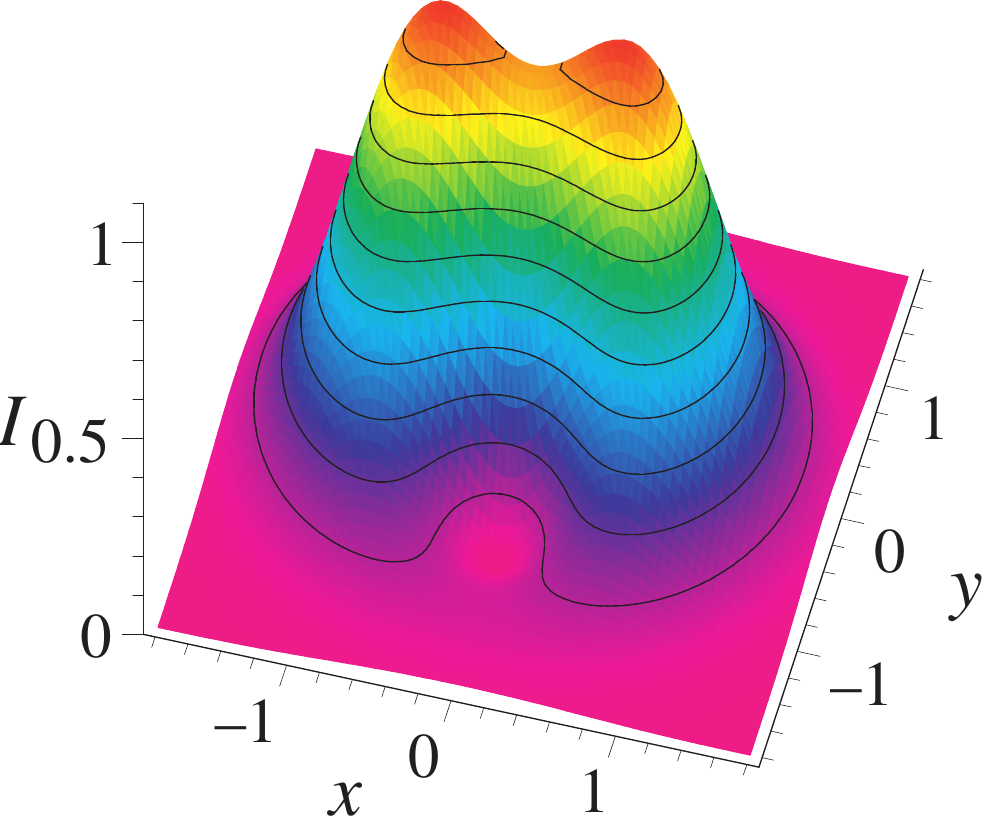}
     \put(-110,105){\rotatebox{0}{\mbox{(a)}}}
  \end{minipage}   % no empty line here so the figures are side by side
  \hspace{0.035\linewidth}
  \begin{minipage}[b]{0.47\linewidth}
     \includegraphics[width=0.71\linewidth,height=0.7\linewidth]{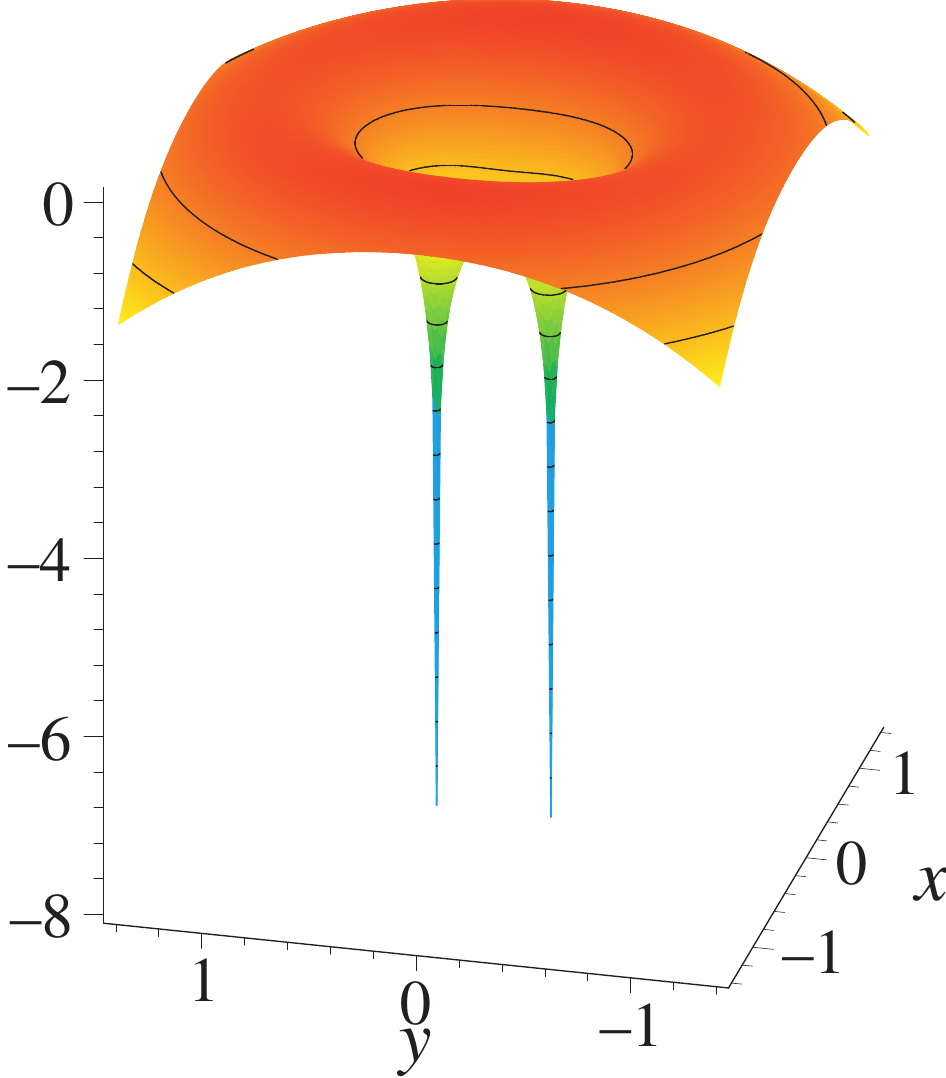}
     \put(-140,30){\rotatebox{90}{\mbox{$\log_{10}(I(x,y,0))$}}}
     \put(-130,105){\rotatebox{0}{\mbox{(b)}}}
  \end{minipage}
\end{center}
\caption{ The focal intensity distribution of superposition $
  u_{0,2}(x,y,z) + u_{0,0}(x,y,-z)$ {(a)} and its logarithm for
  superposition $u_{0,2}(x,y,z) + \frac{1}{10} u_{0,0}(x,y,-z)$ {(b)}
  demonstrate that narrowly wound bright helices self-overlap and
  become ill defined whereas dark helices remain distinguishable [coordinate 
  axes scaled in units of beam width~$w_0$].}
\label{Fig_3_self-overlapping_double_helices}
\end{figure}

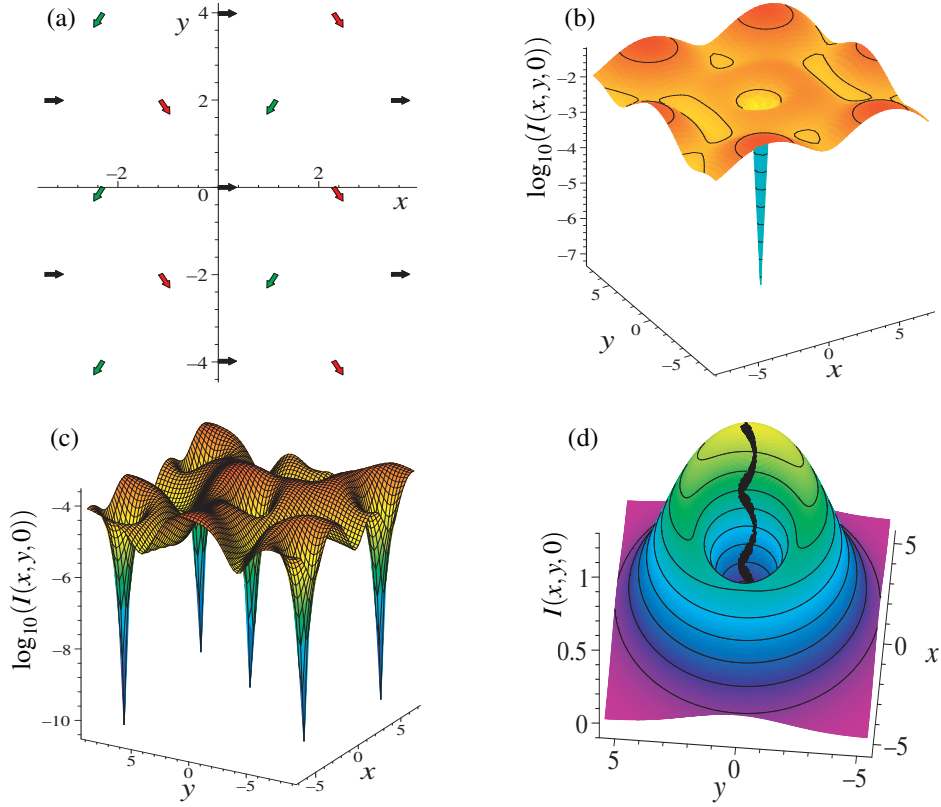
\begin{figure*}[h]%[!t]
\begin{center}
  \begin{minipage}[b]{0.475\linewidth} % "b" captions on the same line
     \includegraphics[width=0.8\linewidth,height=0.8\linewidth]{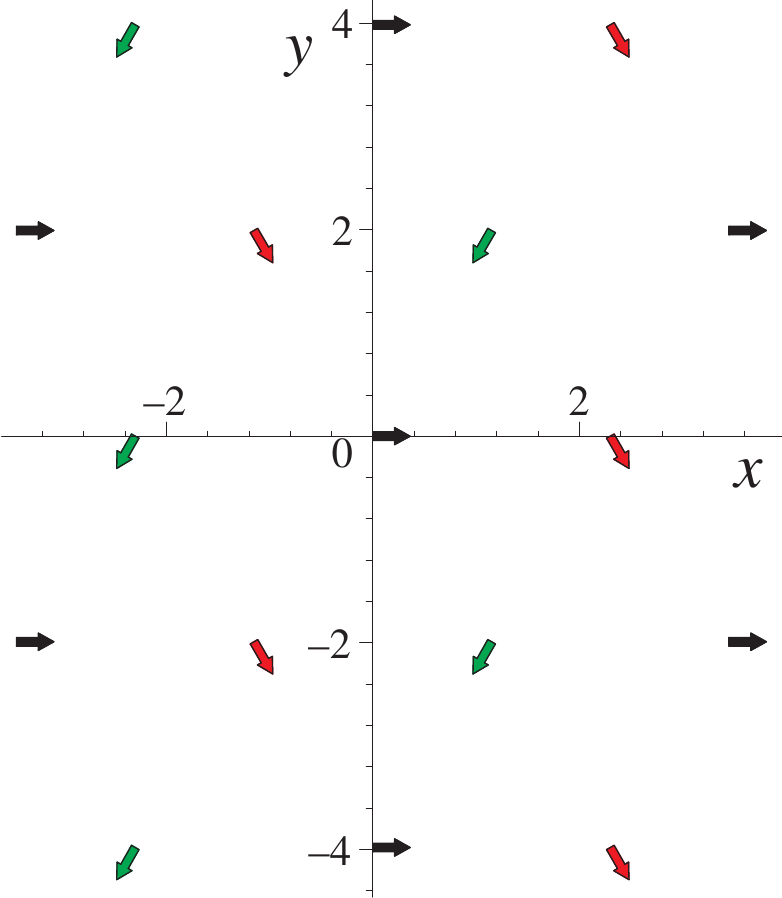}
     \put(-140,135){\rotatebox{0}{\mbox{(a)}}}
   \end{minipage} % no empty line here so the figures are side by side
  \hspace{0.03\linewidth}
  \begin{minipage}[b]{0.475\linewidth}
     \includegraphics[width=0.8\linewidth,height=0.8\linewidth]{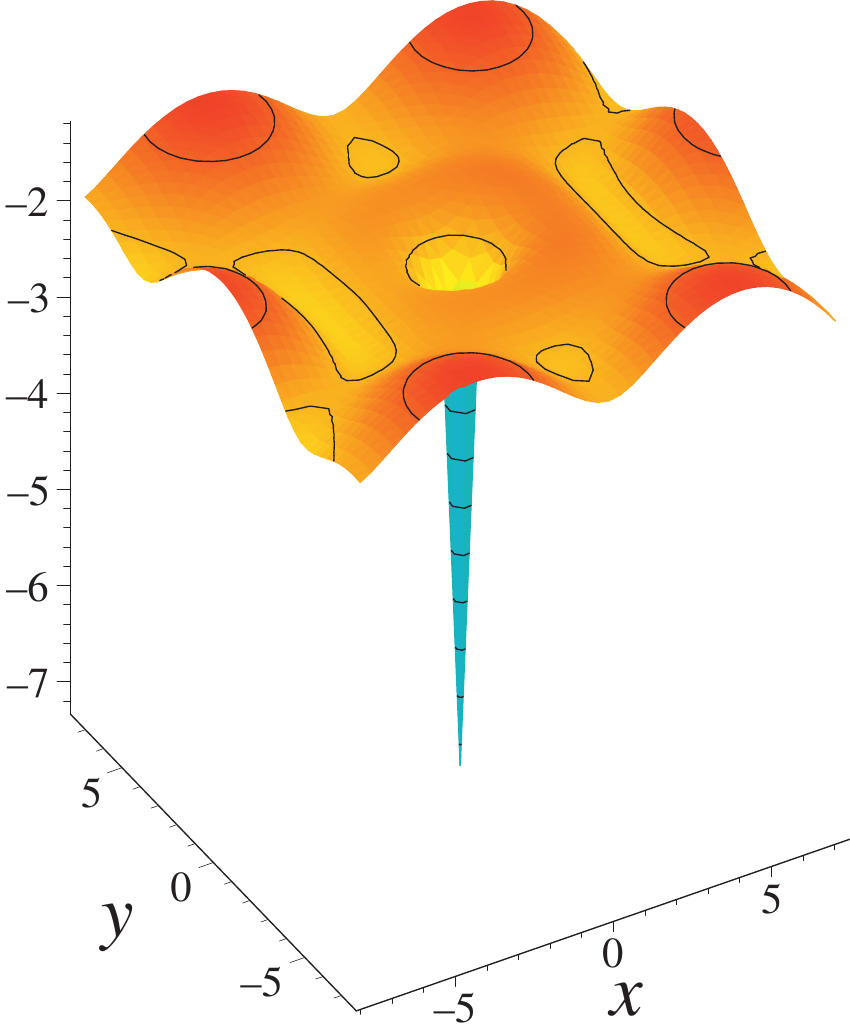}
     \put(-155,70){\rotatebox{90}{\mbox{$\log_{10}(I(x,y,0))$}}}
     \put(-140,135){\rotatebox{0}{\mbox{(b)}}}
  \end{minipage}
  \hspace{0.03\linewidth}
\\
\vspace{0.5cm}
  \begin{minipage}[b]{0.475\linewidth}
    \includegraphics[width=0.8\linewidth,height=0.8\linewidth]{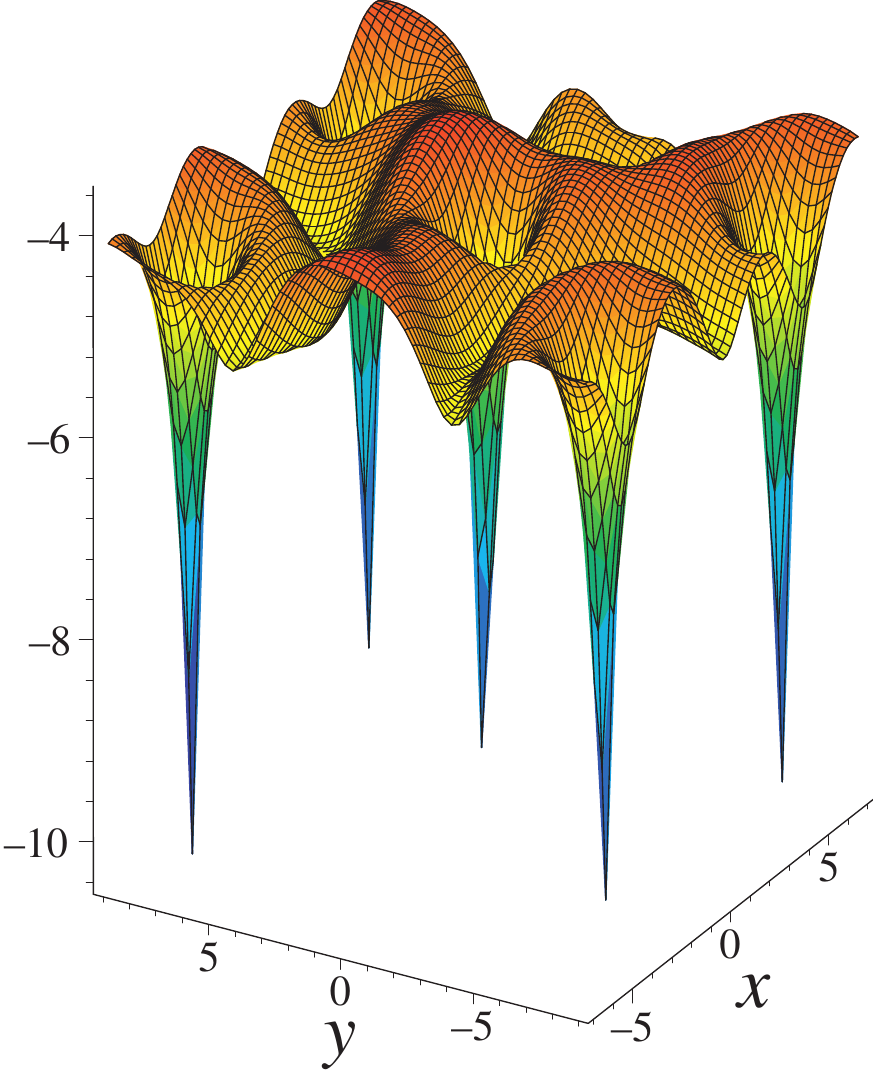}
     \put(-155,50){\rotatebox{90}{\mbox{$\log_{10}(I(x,y,0))$}}}
     \put(-140,135){\rotatebox{0}{\mbox{(c)}}}
  \end{minipage}
  \hspace{0.03\linewidth}
  \begin{minipage}[b]{0.475\linewidth}
     \includegraphics[width=0.8\linewidth,height=0.8\linewidth]{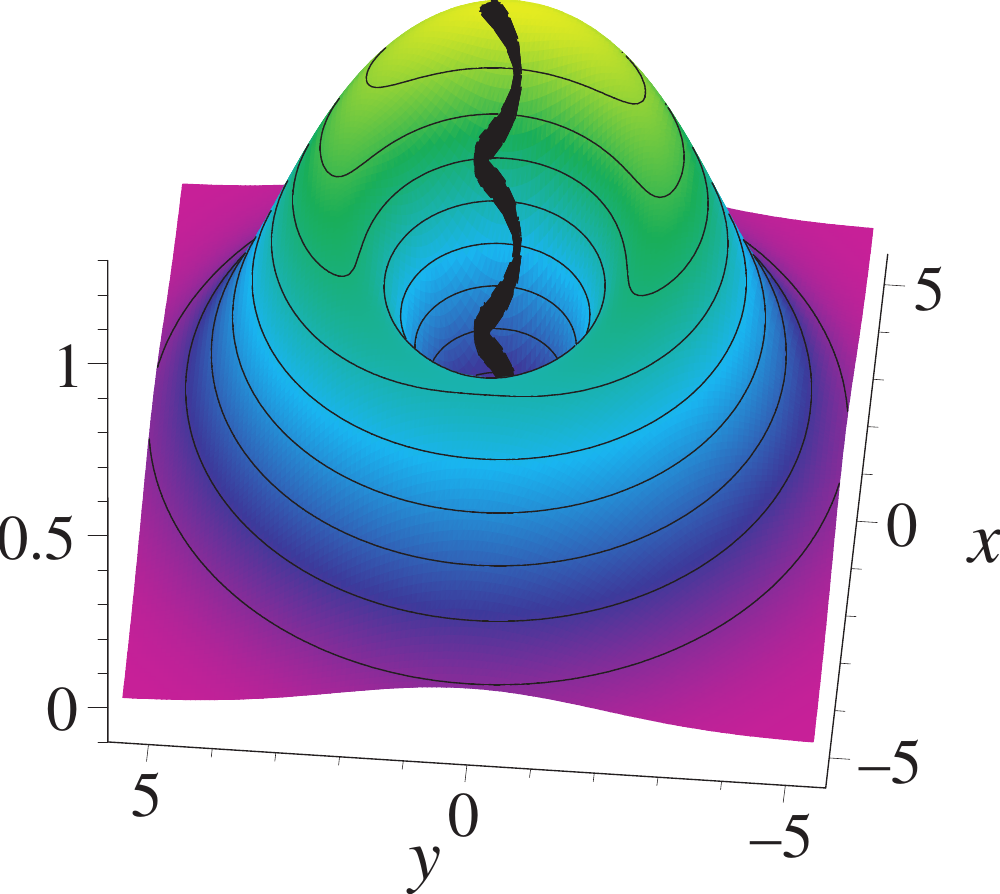}
     \put(-150,70){\rotatebox{90}{\mbox{$I(x,y,0)$}}}
     \put(-140,135){\rotatebox{0}{\mbox{(d)}}}
  \end{minipage}
\end{center}
\caption{A set of parallel equal Gaussian beams
  $u_{0,0}(x-X_G,y-Y_G,z)$ travelling along the $z$-axis and with
  their beam axes centered on a set of grid points $\{(X_G,Y_G,0)\}$
  of a hexagonal lattice is arranged into three sub-lattices with
  different polarization orientations (beam centers and respective
  beam polarizations indicated by arrows in panel~{(a)}; nearest
  neighbour distance $2.3\, w_0$). Despite its tight packing this
  sub-lattice arrangement avoids destructive interference between
  beams of equal polarization and yields a roughly uniformly bright
  background, see panels~{(b)} and~{(c)}. The focal intensity
  distribution of a single dark helix created from the superposition
  $\Sigma(x,y,z;\frac{1}{8})$ of eq.~(\ref{eq_single_helix_Sigma}) is
  displayed in panel~{(d)} together with an artist's impression of the
  location of its dark helix as a black line. Such dark helices can be
  inserted into the bright background without losing their
  contrast,~{(b)} and~{(c)}, for details see
  section~\ref{sec_4_LatticesOfDarkHelices}. Massively parallel or
  sparse implementation of dark helices with good contrast is
  possible. All position coordinate axes are scaled in units of beam
  width~$w_0$.}
\label{Fig_4_GridOfSpirals}
\end{figure*}

Unlike bright features dark features can provide
superresolution~\cite{Hell_SCI07}. This allows us to create structures
with spatial extensions below the diffraction limit. To see how the
optical resolution limit constrains bright helices much more than dark
ones, recall that bright helices are centered on the maximum
position~$\mathbf{r_M}$ of the local electric field. According to
Fourier's theorems monochromatic light fields in free space have to be
smooth and differentiable, they cannot have spikes. Therefore, the
\emph{field's magnitude}, when moving by a small displacement $\delta
\mathbf{r}$ away from the core of the helix, can at best drop
quadratically with $\delta \mathbf{r}$, namely, with~$\kappa$ as the
maximal local curvature we have
\begin{eqnarray}%{l}
\label{eq_local_trends_maxima}
|\delta \mathbf{E_M}(\delta \mathbf{r})| =
|\mathbf{E}(\mathbf{r_M})| - | \mathbf{E}(\mathbf{r_M} + \delta \mathbf{r})| 
\leq \frac{|\kappa|}{2} \cdot (\delta \mathbf{r})^2 + 
\mathbf{\it O}(( \delta \mathbf{r})^{3}) \; .
\end{eqnarray}
Around a zero at location~$\mathbf{r_m}$ the field instead typically
rises linearly (unless one tracks the helix core), this variation is
bounded by the largest local gradient~$\gamma$
\begin{eqnarray}%{l}
\label{eq_local_trends_minima}
|\delta \mathbf{E_m}(\delta \mathbf{r})| \approx
| \mathbf{E}(\mathbf{r_m} + \delta \mathbf{r})| - 0 %|\mathbf{E}(\mathbf{r_M})| 
\leq |\gamma|  \cdot |\delta \mathbf{r}|  + \mathbf{\it O}(( \delta \mathbf{r})^{2}) \; .
\end{eqnarray}
This difference in the scaling of the variation of the local field
magnitude allows for sharper definition of dark helices compared with
bright ones; this is represented graphically in the field magnitude
plot of panel (c) of Fig.~\ref{Fig_1_single_helix}, in the comparison
in Fig.~\ref{Fig_3_self-overlapping_double_helices}, and in the
`spikiness' of the dark helices' intensity profiles in
Fig.~\ref{Fig_4_GridOfSpirals}~{(b)} and~{(c)}. Note that the location
of the bright helices in these two plots are harder to make out than
those of the dark ones. One might be tempted to argue that a
logarithmic plot `unfairly' pronounces the presence of dark helices
over bright ones, but photoresists, for example, tend to have
logarithmic response functions~\cite{Kim_OE11} so the use of
Figs.~\ref{Fig_3_self-overlapping_double_helices}~{(b)},
and~\ref{Fig_4_GridOfSpirals}~{(b)} and~{(c)} is quite suitable when
trying to gauge outcomes in a photo-lithographic application.

\section{Lattices of dark helices~\label{sec_4_LatticesOfDarkHelices}}

It is possible to embed dark helical beams into a background of
roughly uniform illumination. To this end a hexagonal lattice is used
because it has high symmetry and can be very tightly
packed. Ordinarily, regions of completely destructive interference of
such tightly packed beams would be unavoidable. In order to make
certain that no dark areas other than the desired dark helices are
created, we partition the lattice into three triangular sub-lattices
with different polarization directions, see colour-coding
of~Fig.~\ref{Fig_4_GridOfSpirals}~{(a)}. In our examples, the
hexagonal lattice is either filled with ordinary Gaussian
$u_{0,0}$-beams,~Fig.~\ref{Fig_4_GridOfSpirals} {(b)}, or with
single-helix beams~$\Sigma(C)$,
see~Fig.~\ref{Fig_4_GridOfSpirals}~{(c)}; one of which is displayed
in~Fig.~\ref{Fig_4_GridOfSpirals}~{(d)}.

In case~\ref{Fig_4_GridOfSpirals}~{(b)}, the lattice beam
$u_{0,0}(x,y,z)$ at the origin (black arrow) is removed, a very weak
$+y$-polarized beam is superimposed to remove spill-over from the nearest
neighbours and a single-helix beam~$\Sigma(x,y,z;\frac{1}{8})$
inserted at the origin. This construction demonstrates that a single
dark helix can be embedded into a bright uniform background.
Alternatively, the entire lattice is formed from helical beams with
equal strength, see~Fig.~\ref{Fig_4_GridOfSpirals}~{(c)}; where, on
account of the slightly wider lattice spacing, no spill-over
compensation is applied.

The logarithmic intensity plots~\ref{Fig_4_GridOfSpirals}~{(b)}
and~{(c)} confirm that the background intensity is quite uniform
(varies less than an order of magnitude) and yet allows for embedded
dark spirals with excellent intensity contrast: the helix cores are at
least four orders of magnitude darker than the background.

\section{Applications, conclusions and
outlook~\label{sec_5_ApplicationsConclOutlook}}

The generation of bright and dark intensity helices from paraxial
laser beams was studied. Such intensity helices of light may turn out
to be applied in nanolithography of bulk media~\cite{Kawata_Nat01} to
create sharply contoured helical imprints in potentially massively
parallel fashion.

In section~\ref{sec_dark_scaling} the variation of the
electric field magnitude in the immediate vicinity of intensity
helices was characterized. This variation scales linearly for dark and
quadratically for bright helices. 

The very much sharper intensity contrasts this scaling implies for
dark helices may turn out to be of considerable practical
importance. The logarithmic intensity plots of
Fig.~\ref{Fig_4_GridOfSpirals}~(b) and~(c) show that we can create
sharp spikes of the intensity distribution of dark helices whereas
this is impossible for bright helices. This observation is relevant
for lithographic applications because photoresists tend to show
logarithmic responses to the magnitude of the irradiating
field~\cite{Kim_OE11}.

In order to demonstrate the versatility of the approach presented here
it was shown that single dark helices can be sparsely embedded in a
uniform bright background but also tightly packed into a dense
hexagonal grid embedded in a uniform bright background.

The intensity helices' defined handedness might, for example, help
with the production of chiral optical
meta-materials~\cite{Sihvola_Rev07,Semchenko_EPJAP09}, possibly using
longer wavelengths such as from
THz-radiation~\cite{Tonouchi_NATPhys07}.

Filtering of handed molecules, through a membrane with
photo-lithographically etched helical holes or via
handedness-sensitive trapping in solution, might become possible.

For the manipulation of ultra-cold gas clouds via the optical dipole
force, dark (blue-detuned) and bright
(red-detuned~\cite{Bhattacharya_OC07}) helical beams may turn out to
be useful~\cite{Chu_RMP98}. 

The fact that helices can be made to intertwine, see
section~\ref{sec_2_interference}, opens up the possibility of studying
intertwined transport. In the presence of gravity or other uniform
force fields a helix that is tilted away from vertical by more than
its pitch angle can potentially serve as a micros\-copic
Archimedean-screw transporter~\cite{Kuhr_SCI95} or a helical
ratchet-potential. Transport along helices is expected to show
localization behaviour~\cite{Gaididei_NJP05,Exner_PLA07} and other
interesting quantum-transport phenomena~\cite{Qi_PRB09}.

\section*{Acknowledgment}
I would like to thank Paul Kaye for proof-reading of this manuscript.


\begin{thebibliography}{10}
\newcommand{\enquote}[1]{``#1''}

\bibitem{Allen_PRA92}
L.~{Allen}, M.~W. {Beijersbergen}, R.~J.~C. {Spreeuw}, and J.~P. {Woerdman},
  \enquote{{Orbital angular momentum of light and the transformation of
  Laguerre-Gaussian laser modes},} Phys. Rev. A \textbf{45}, 8185--8189 (1992).

\bibitem{Padgett_CP00}
M.~{Padgett} and L.~{Allen}, \enquote{{Light with a twist in its tail},} Cont.
  Phys. \textbf{41}, 275--285 (2000).

\bibitem{Padgett_PT04}
M.~{Padgett}, J.~{Courtial}, and L.~{Allen}, \enquote{{Light's Orbital Angular
  Momentum},} Physics Today \textbf{57}, 35--40 (2004).

\bibitem{Harris_OC94}
M.~{Harris}, C.~A. {Hill}, and J.~M. {Vaughan}, \enquote{{Optical helices and
  spiral interference fringes},} Opt. Commun. \textbf{106}, 161--166 (1994).

\bibitem{Vaughan_JOPA99}
J.~M. {Vaughan}, \enquote{{Interferometry, atoms and light scattering: one
  hundred years of optics},} J. Opt. A: Pure Appl. Opt. \textbf{1}, 750--768
  (1999).

\bibitem{Leach_OE06}
J.~{Leach}, S.~{Keen}, M.~J. {Padgett}, C.~{Saunter}, and G.~D. {Love},
  \enquote{{Direct measurement of the skew angle of the Poynting vector in a
  helically phased beam},} Optics Express \textbf{14}, 11919--11924 (2006).

\bibitem{Ole_AJP05}
O.~{Steuernagel}, \enquote{{Equivalence between focused paraxial beams and the
  quantum harmonic oscillator},} Am. J. Phys. \textbf{73}, 625--629 (2005).

\bibitem{Lekner_JOPA04}
J.~{Lekner}, \enquote{{LETTER TO THE EDITOR: Helical light pulses},} J. Opt. A:
  Pure Appl. Opt. \textbf{6}, L29--L32 (2004).

\bibitem{Hamazaki_OE06}
J.~{Hamazaki}, Y.~{Mineta}, K.~{Oka}, and R.~{Morita}, \enquote{{Direct
  observation of Gouy phase shift in a propagating optical vortex},} Opt.
  Express \textbf{14}, 8382--8392 (2006).

\bibitem{Baumann_OE09}
S.~M. {Baumann}, D.~M. {Kalb}, L.~H. {MacMillan}, and E.~J. {Galvez},
  \enquote{{Propagation dynamics of optical vortices due to Gouy phase},} Opt.
  Express \textbf{17}, 9818--9827 (2009).

\bibitem{Becker_OE11}
J.~Becker, P.~Rose, M.~Boguslawski, and C.~Denz, \enquote{Systematic approach
  to complex periodic vortex and helix lattices,} Opt. Express \textbf{19},
  9848--9862 (2011).

\bibitem{Volke-SepulvedaJPB09}
K.~{Volke-Sep{\'u}lveda} and R.~{J{\'a}uregui}, \enquote{{All-optical 3D atomic
  loops generated with Bessel light fields},} J. Phys. B: At. Mol. Phys.
  \textbf{42}, 085303 (2009).

\bibitem{Staliunas_cond.mat.99}
K.~{Staliunas}, \enquote{{Vortex Creation in Bose-Einstein Condensates by
  Diffraction on a Helical Light Grating},}
  http://www.arxiv.org/abs/cond-mat/9912268  (1999).

\bibitem{Bhattacharya_OC07}
M.~{Bhattacharya}, \enquote{{Lattice with a twist: Helical waveguides for
  ultracold matter},} Opt. Commun. \textbf{279}, 219--222 (2007).

\bibitem{Nye_Berry_PRSLA74}
J.~F. {Nye} and M.~V. {Berry}, \enquote{{Dislocations in Wave Trains},} Royal
  Society of London Proceedings Series A \textbf{336}, 165--190 (1974).

\bibitem{Kuhr_SCI95}
S.~{Kuhr}, W.~{Alt}, D.~{Schrader}, M.~{M{\"u}ller}, V.~{Gomer}, and
  D.~{Meschede}, \enquote{{Deterministic Delivery of a Single Atom},} Science
  \textbf{293}, 278--281 (2001).

\bibitem{Meschede_Metcalf_03}
D.~Meschede and H.~Metcalf, \enquote{Atomic nanofabrication: atomic deposition
  and lithography by laser and magnetic forces,} J. Phys. D: Appl. Phys.
  \textbf{36}, R17--R38 (2003).

\bibitem{Marrucci_PRL06}
L.~Marrucci, C.~Manzo, and D.~Paparo, \enquote{Optical spin-to-orbital angular
  momentum conversion in inhomogeneous anisotropic media,} Phys. Rev. Lett.
  \textbf{96}, 163905 (2006).

\bibitem{Leach_NAT04}
J.~{Leach}, M.~R. {Dennis}, J.~{Courtial}, and M.~J. {Padgett}, \enquote{{Laser
  beams: Knotted threads of darkness},} Nature \textbf{432}, 165--165 (2004).

\bibitem{MacDonald_SCI02}
M.~P. {MacDonald}, L.~{Paterson}, K.~{Volke-Sepulveda}, J.~{Arlt},
  W.~{Sibbett}, and K.~{Dholakia}, \enquote{{Creation and Manipulation of
  Three-Dimensional Optically Trapped Structures},} Science \textbf{296},
  1101--1103 (2002).

\bibitem{Pampaloni_2004}
F.~Pampaloni and J.~Enderlein, \enquote{{Gaussian, Hermite-Gaussian, and
  Laguerre-Gaussian beams: A primer },}
  http://www.arxiv.org/abs/physics/0410021  (2004).

\bibitem{Haus.book}
H.~A. Haus, \emph{Electromagnetic Noise and Quantum Optical Measurements}
  (Springer, Heidelberg, 2000).

\bibitem{Paterson_SCI01}
L.~{Paterson}, M.~P. {MacDonald}, J.~{Arlt}, W.~{Sibbett}, P.~E. {Bryant}, and
  K.~{Dholakia}, \enquote{{Controlled Rotation of Optically Trapped Microscopic
  Particles},} Science \textbf{292}, 912--914 (2001).

\bibitem{Dholakia_PhysWorld02}
K.~{Dholakia}, G.~C. {Spalding}, and M.~{MacDonald}, \enquote{{Optical
  tweezers: The next generation},} Phys.World \textbf{15}, 31--35 (2002).

\bibitem{MacDonald_OC02}
M.~P. {MacDonald}, K.~{Volke-Sepulveda}, L.~{Paterson}, J.~{Arlt},
  W.~{Sibbett}, and K.~{Dholakia}, \enquote{{Revolving interference patterns
  for the rotation of optically trapped particles},} Opt. Commun. \textbf{201},
  21--28 (2002).

\bibitem{Chu_RMP98}
S.~{Chu}, \enquote{{Nobel Lecture: The manipulation of neutral particles},}
  Rev. Mod. Phys. \textbf{70}, 685--706 (1998).

\bibitem{Hell_SCI07}
S.~W. {Hell}, \enquote{{Far-Field Optical Nanoscopy},} Science \textbf{316},
  1153--1158 (2007).

\bibitem{Kim_OE11}
Y.~{Kim}, H.~{Jung}, S.~{Kim}, J.~{Jang}, J.~Y. {Lee}, and J.~W. {Hahn},
  \enquote{{Accurate near-field lithography modeling and quantitative mapping
  of the near-field distribution of a plasmonic nanoaperture in a metal},} Opt.
  Express \textbf{19}, 19296--19309 (2011).

\bibitem{Kawata_Nat01}
S.~{Kawata}, H.-B. {Sun}, T.~{Tanaka}, and K.~{Takada}, \enquote{{Finer
  features for functional microdevices},} Nature \textbf{412}, 697--698 (2001).

\bibitem{Sihvola_Rev07}
A.~{Sihvola}, \enquote{{Metamaterials in electromagnetics},} Metamaterials
  \textbf{1}, 2--11 (2007).

\bibitem{Semchenko_EPJAP09}
I.~V. {Semchenko}, S.~A. {Khakhomov}, and S.~A. {Tretyakov}, \enquote{{Chiral
  metamaterial with unit negative refraction index},} Eur. Phys. J. Appl. Phys.
  \textbf{46}, 032607 (2009).

\bibitem{Tonouchi_NATPhys07}
M.~{Tonouchi}, \enquote{{Cutting-edge terahertz technology},} Nature Photonics
  \textbf{1}, 97--105 (2007).

\bibitem{Gaididei_NJP05}
Y.~B. {Gaididei}, P.~L. {Christiansen}, P.~G. {Kevrekidis}, H.~{B{\"u}ttner},
  and A.~R. {Bishop}, \enquote{{Localization of nonlinear excitations in curved
  waveguides},} New J. Phys. \textbf{7}, 52--52 (2005).

\bibitem{Exner_PLA07}
P.~{Exner} and M.~{Fraas}, \enquote{{A remark on helical waveguides},} Phys.
  Lett. A \textbf{369}, 393--399 (2007).

\bibitem{Qi_PRB09}
X.-L. {Qi} and S.-C. {Zhang}, \enquote{{Field-induced gap and quantized charge
  pumping in a nanoscale helical wire},} \prb \textbf{79}, 235442 (2009).

\end{thebibliography}
\end{document}